\documentclass[sigconf]{acmart}

\usepackage{multirow}
\usepackage{subcaption}
\usepackage{caption}
\AtBeginDocument{%
  }

\setcopyright{acmlicensed}
\copyrightyear{2018}
\acmYear{2018}
\acmDOI{XXXXXXX.XXXXXXX}
\acmConference[Conference acronym 'XX]{Make sure to enter the correct
  conference title from your rights confirmation email}{June 03--05,
  2018}{Woodstock, NY}
\acmISBN{978-1-4503-XXXX-X/2018/06}

\begin{document}

\title{HRFT: Mining High-Frequency Risk Factor Collections End-to-End via Transformer}

\author{Wenyan Xu}
\affiliation{
    \institution{School of Statistics and Mathematics, Central University of Finance and Economics}
    \city{Beijing}
    \country{China}
}
\email{2022211032@email.cufe.edu.cn}

\author{Rundong Wang}
\affiliation{
    \institution{TiMi Studio, Tencent}
    \city{Chengdu}
    \country{China}
}
\email{rundongwang@tencent.com}

\author{Chen Li}
\authornote{Corresponding authors.}
\affiliation{
    \institution{Computer Network Information Center, Chinese Academy of Sciences}
    \city{Beijing}
    \country{China}
}
\email{lichen@sccas.cn}

\author{Yonghong Hu}
\affiliation{
    \institution{School of Statistics and Mathematics, Central University of Finance and Economics}
    \city{Beijing}
    \country{China}
}
\email{huyonghong@cufe.edu.cn}

\author{Zhonghua Lu}
\authornotemark[1]
\affiliation{
    \institution{Computer Network Information Center, Chinese Academy of Sciences}
    \city{Beijing}
    \country{China}
}
\email{zhlu@sccas.cn}

\renewcommand{\shortauthors}{Wenyan Xu et al.}

\copyrightyear{2025}
\acmYear{2025}
\setcopyright{cc}
\setcctype{by}
\acmConference[WWW Companion '25]{Companion Proceedings of the ACM Web Conference 2025}{April 28-May 2, 2025}{Sydney, NSW, Australia}
\acmBooktitle{Companion Proceedings of the ACM Web Conference 2025 (WWW Companion '25), April 28-May 2, 2025, Sydney, NSW, Australia}
\acmDOI{10.1145/3701716.3715235}
\acmISBN{979-8-4007-1331-6/25/04}

% \copyrightyear{2025}
% \acmYear{2025}
% \setcopyright{acmlicensed}
% \acmConference[WWW Companion '25] {Companion of the 16th ACM/SPEC International Conference on Performance Engineering}{April 28-May 2, 2025}{Sydney, NSW, Australia.}
% \acmBooktitle{Companion of the 16th ACM/SPEC International Conference on Performance Engineering (WWW Companion '25), April 28-May 2, 2025, Sydney, NSW, Australia}
% \acmISBN{979-8-4007-1331-6/25/04}
% \acmDOI{10.1145/XXXXXX.XXXXXX}

\begin{abstract}
In quantitative trading, transforming historical stock data into interpretable, formulaic risk factors enhances the identification of market volatility and risk. Despite recent advancements in neural networks for extracting latent risk factors, these models remain limited to feature extraction and lack explicit, formulaic risk factor designs. By viewing symbolic mathematics as a language—where valid mathematical expressions serve as meaningful "sentences"—we propose framing the task of mining formulaic risk factors as a language modeling problem. In this paper, we introduce an end-to-end methodology, Intraday Risk Factor Transformer (IRFT), to directly generate complete formulaic risk factors, including constants. We use a hybrid symbolic-numeric vocabulary where symbolic tokens represent operators and stock features, and numeric tokens represent constants. We train a Transformer model on high-frequency trading (HFT) datasets to generate risk factors without relying on a predefined skeleton of operators (e.g., $+, \times, /, \sqrt{x}, \log{x}, \cos{x}$). 
It determines the general form of the stock volatility law, including constants; for example, $f(x) = \tan(ax + b)$, where $x$ is the stock price. We refine the predicted constants ($a, b$) using the Broyden--Fletcher--Goldfarb--Shanno (BFGS) algorithm to mitigate non-linear issues. Compared to the ten approaches in SRBench, an active benchmark for symbolic regression (SR), IRFT achieves a 30\% higher investment return on the HS300 and S\&P500 datasets, while achieving inference times that are orders of magnitude faster than existing methods in HF risk factor mining tasks. Our code and dataset are publicly accessible via the following GitHub repository: \url{https://github.com/wencyxu/IRF-LLM-accepted-at-WWW25-}.
\end{abstract}




\maketitle

\section{Introduction}
Most risk factor mining traditionally relies on factor models and requires the manual screening of covariates, also known as risk factors, such as beta \cite{prices1964theory}, size and value \cite{fama1992cross}, momentum \cite{carhart1997persistence}, and residual volatility and liquidity \cite{sheikh1996barra}. However, these hand-picked factors often show weak correlations with stock return volatility and fail to keep pace with market dynamics. Statistical models such as Principal Component Analysis (PCA) \cite{fan2021augmented} and factor analysis \cite{ang2020using} can identify latent risk factors but are limited in capturing nonlinear risks. Deep Risk Model (DRM) \cite{lin2021deep} address this limitation by utilizing neural networks to learn and extract latent risk factors directly from stock data, replacing the traditional manually designed covariance matrices. However, when it comes to designing explicit signals—formulaic risk factors representing market trends—neural networks are relatively limited, being primarily restricted to feature extraction.

Symbolic mathematics can be likened to a language, where mathematically valid expressions function as coherent "sentences." Inspired by this analogy, we frame the task of extracting formulaic risk factors as a language modeling problem. Thus, employing deep language models to process symbolic mathematics emerges as a natural approach. Inspired by symbolic regression (SR)—which focuses on discovering general mathematical relationships from large datasets \cite{becker2023predicting,xu2025mining}—we aim to uncover patterns underlying the volatility in historical stock data.

 We generate formulaic HF risk factors from raw stock trading features to measure short-term market volatility for the first time. These HF risk factors effectively measure the variance of future stock return distributions. We train a Transformer model from scratch on the HFT dataset instead of fine-tuning an open-source Pre-Trained Model (PTM). Note that most of the capabilities of language models lie in semantic processing. However, the task of generating formulas primarily focuses on tackling numbers and symbols, rather than comprehending extensive vocabularies. So to represent formula factors as sequences, we use direct Polish notation. Operators, variables, and integers are single autonomous tokens. Constants are represented as sequences of 3 tokens: sign, mantissa, and exponent. We preprocess open/close/high/low/volume/vwap features using word expressions. Our embedder and transformer learn from word expressions of stock features as input and formulaic factors as output. The sequence-to-sequence Transformer architecture has 16 attention heads and an embedding dimension of 512, totaling 86M parameters.

The main contributions are summarized as follows: 
\begin{itemize}
\item We propose a data generator for end-to-end HF risk factor mining for downstream tasks and short-term stock market risk measurements.
\item To do the HF risk factor mining, we employ a hybrid symbolic-numeric vocabulary where symbolic tokens represent operat\\ors$/$stock features and numeric tokens represent constants, then train a Transformer model over the HFT dataset to directly generate full formula factors without relying on skeletons. We refine the prediction constants during the Broyden–Fletcher–Goldfarb–Shanno optimization process (BFGS) as informed guessing, which can alleviate the non-linear issues effectively. Lastly, we utilize generation and inference techniques that enable our model to scale to complex realistic datasets, whereas current works are only suitable for synthetic datasets. 
\item 
Our method outperforms the 10 approaches in SRBench in investment simulation experiments over the
HS300 and S\&P500 dataset, achieving a 30\% higher investment return (refer to Fig. \ref{fig:heatmap_sp500}(b)) and several orders of magnitude faster inference time in factor mining tasks (refer to Fig. \ref{fig:pareto}). 
\end{itemize}

\begin{figure}[H]
  \centering \includegraphics[width=0.5\linewidth]{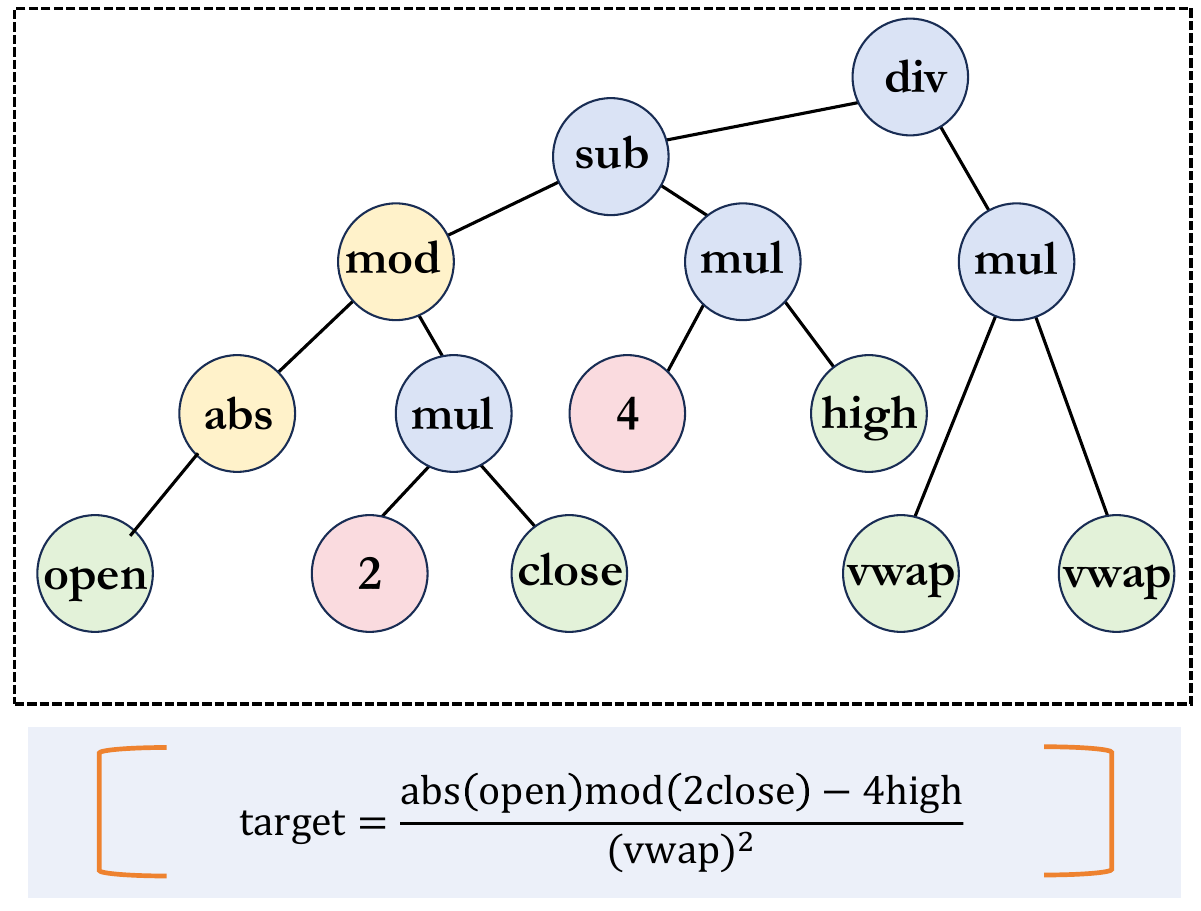}
  \caption{A formulaic HF risk factor can be represented as a binary tree. The nodes are unary operators like ‘mod’ and ‘abs’, and binary operators like ‘sub’, ‘div’, and ‘mul’. The leaves are input features such as ‘high’, ‘open’, ‘close’, and ‘vwap’, along with constants like 2 and 4.}
  \label{fig:tree}
\end{figure}

\begin{table*}
\centering
  \caption{Detailed parameter settings of IRFT.}
  
  \label{tab:parameters}
  \begin{tabular}{lcccc}
    \toprule
    Parameter & Description & \multicolumn{2}{c}{Value} \\
    \cmidrule(r){1-2} \cmidrule(l){3-4}
    & & W=5(S\&P500) & W=5(SSE50) \\
    \midrule
    W\_max & Max input dimension & \multicolumn{2}{c}{10} \\
    D\_((a,b)) & Distribution of (a, b) in an affine transformation & \multicolumn{2}{c}{sign~U{-1,1}, mantissa~U(0,1), exponent~U(-2,2)} \\
    b\_max & Max binary operators & 5+W & 5+W \\
    O\_b & Binary operators & \multicolumn{2}{c}{add,sub,mul,div} \\
    u\_max & Max unary operators & 5 & 5 \\
    O\_u & Unary operators & \multicolumn{2}{c}{inv,abs,sqr,sqrt,sin,cos,tan,atan,log,exp} \\
    M\_min & Min number of points & \multicolumn{2}{c}{10D} \\
    M\_max & Max number of points & \multicolumn{2}{c}{200} \\
    c\_max & Max number of candidate function clusters & \multicolumn{2}{c}{10} \\
    \bottomrule
  \end{tabular}
\end{table*}

\section{Data generation}
Instead of using an existing PTM, we provide one. We train a pre-trained language model on HFT datasets. Each training sample $(x,y) \in \mathbb{R}^W \times \mathbb{R}$  is input in pairs, aiming to generate an HF risk factor expression $E$ that satisfies $y = E(x)$. For example, we first randomly sample an HF risk factor expression $E$, then sample a set of $M$ input values in $\mathbb{R}^W$, $\{x_k \mid k \in 1, \ldots, M\}$, and calculate $y_k = E(x_k)$.

\subsection{Generating functions}
Inspired by \cite{lample2019deep}, we randomly generate tree structures composed of mathematical operators, variables, and constants, as shown in Fig. \ref{fig:tree}. 
\begin{enumerate}
    \item Sample an integer $W$ from $U\{1, W_{\max}\}$, as the expected input dimension of the function $f$.
    \item Sample an integer $b$ from $U\{W-1, W+b_{\max}\}$, as the number of binary operators, and then randomly select $b$ binary operators from the set $U\{\text{add}, \text{sub}, \text{mul}, \text{div}\}$. Note that all the binary/unary operators are shown in the sets $O_b/O_u$. (see Table \ref{tab:operators}).
    \item Construct a binary tree   using these $b$ binary operator, drawing on the method of \cite{lample2019deep}.
    \item For each leaf node in the tree, randomly select one of the input HF features $x_w$, where $w \in 1, \ldots, W$. HFT datasets can be classified into two types: the China and the U.S. stock market.
    \item Sample an integer $u$ from $U\{0, u_{\max}\}$, as the number of unary operators, and then randomly select $u$ unary operators from $O_u$ in table \ref{tab:operators}, and insert them randomly into any position in the tree.
    \item For each variable $x_w$ and the unary operator $u$, apply the random affine transformation, that is, replace $x_w$ with $ax_w + b$, and replace $u$ with $au + b$, where $(a,b)$ follows the distribution $D_{(a,b)}$.

\end{enumerate}
To independently control the number of unary operators (unrelated to $W$) and binary operators (related to $W$), we cannot directly sample a unary-binary tree as done by \cite{lample2019deep}. We ensure the first $W$ variables are present in the tree, avoiding functions like $x_2 + x_4$ that lack intermediate variables. To enhance expression diversity, we randomly drop out HF variables, allowing our model to set the coefficient of $x_3$ to zero. By randomly sampling tree structures and numerical constants, our model rarely encounters identical functions, preventing simple memorization. Detailed parameters for sampling HFT risk factor expressions are in Table \ref{tab:parameters}. 

Table \ref{tab:parameters} shows the detailed parameter settings of our data generator when using the IRFT model to sample HF risk factor expressions on two types of HFT data sets in the China and the U.S. stock market. Among them, $W_{\max}$ represents the maximum feature dimension of the input data set allowed by our model, which is 10. $M_{\min}/M_{\max}$ represents the minimum/maximum value of the logarithm of the sample pairs contained in a data packet $B$. $b_{\max}/u_{\max}$ represents the number of binary/unary operators sampled. $O_b/O_u$ represents the set of binary/unary operators sampled. $D_{(a,b)}$ represents the coefficient distribution of random affine transformation for each variable $x_w$ and unary operator $u$, that is, replacing $x$ with $ax_w+b$, replacing $u$ with $au+b$; $c_{\max}$ is the number of candidate functions in the refine stage.

\subsection{Generating inputs} \label{Generating inputs}
Before generating the HF risk factor $E: \mathbb{R}^W \to \mathbb{R}$, we first sample $M \in U\{10W, M_{\max}\}$ HF feature values $x_k \in \mathbb{R}^W$ from the distribution $D_x$. Then, we calculate their RV values $y_k = E(x_k)$, and feed them into IRFT. If $x_k$ is not in the domain of $E$ or if $y_k$ exceeds $10^{100}$, IRFT model will terminate the current generation process and initiate the generation of a new HF risk factor $E$. This approach aims to focus the model's learning as much as possible on the domain of $E$, providing a simple and effective method. To enhance the diversity of the input distribution during training, we select input samples from a mixture distribution composed of $c$ random centers. The following steps illustrate this process, using distribution $D=2$ as an example (Gaussian or uniform):
\begin{enumerate}
    \item Randomly sample $c \sim U\{1, c_{\max}\}$ clusters. Each cluster has a weight $w_i \sim U(0,1)$, such that $\sum_i w_i = 1$.
    \item For each cluster $i \in N_i$, randomly draw a centroid $\mu_i \sim \mathcal{N}(0,1)^W$ from a Gaussian distribution, a variance vector $\sigma_i \sim U(0,1)^W$ from a uniform distribution, and a distribution type $D_i \in \{N, U\}$.
    \item For each cluster $i \in N_i$, sample $w_k M$ input points from $D_i (\mu_i, \sigma_i)$, and then apply a random rotation using the Haar distribution. 
    \item Concatenate all the input points and standardize them.
\end{enumerate}

\subsection{Tokenization}
We preprocess HTF stock open/close/high/low/volume/vwap features using word expressions. Our embedder and transformer learn from these feature word expressions as input and formulaic factor word expressions as output. Following \cite{charton2021linear}, we represent numbers in decimal scientific notation with four significant digits, using three tokens: sign, mantissa (0 to 9999), and exponent ($E-100$ to $E100$). Referring to \cite{lample2019deep}, we use prefix expressions (Polish notation) for HF risk factors, encoding operators, variables, and integers, and applying the same method for constants. For example, $f(x) = \tan(9.7341x)$ is encoded as $[tan, mul, +, 97341, E-3, x]$. The Decoder’s vocabulary includes symbolic tokens (operators and variables) and integer tokens, while the Encoder’s vocabulary only includes integer tokens.

\begin{figure*}[h]
  \centering
  \begin{minipage}{0.45\linewidth}  \includegraphics[width=\linewidth]{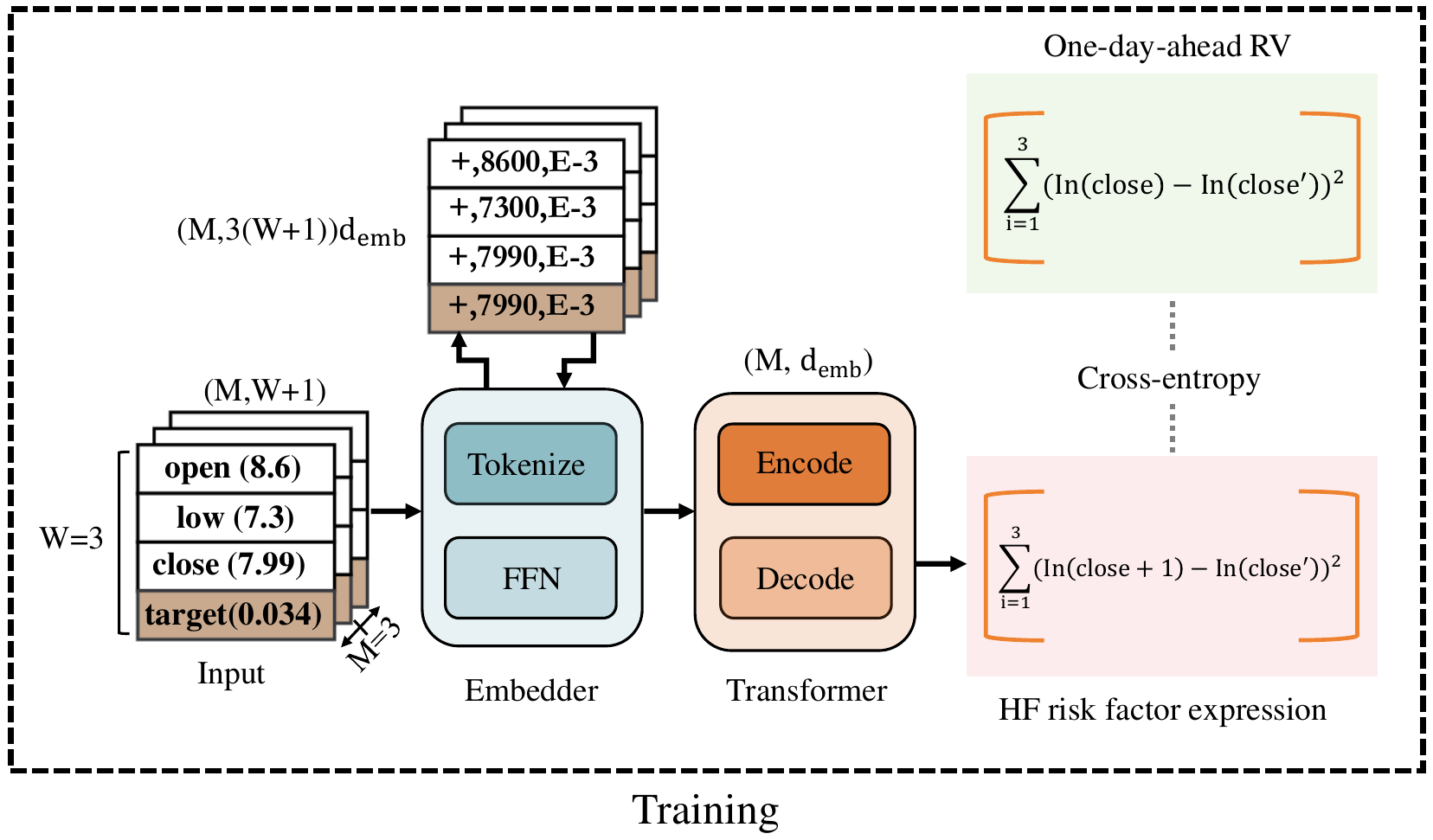}
  \end{minipage}
  \hfill
  \begin{minipage}{0.45\linewidth}
    \includegraphics[width=\linewidth]{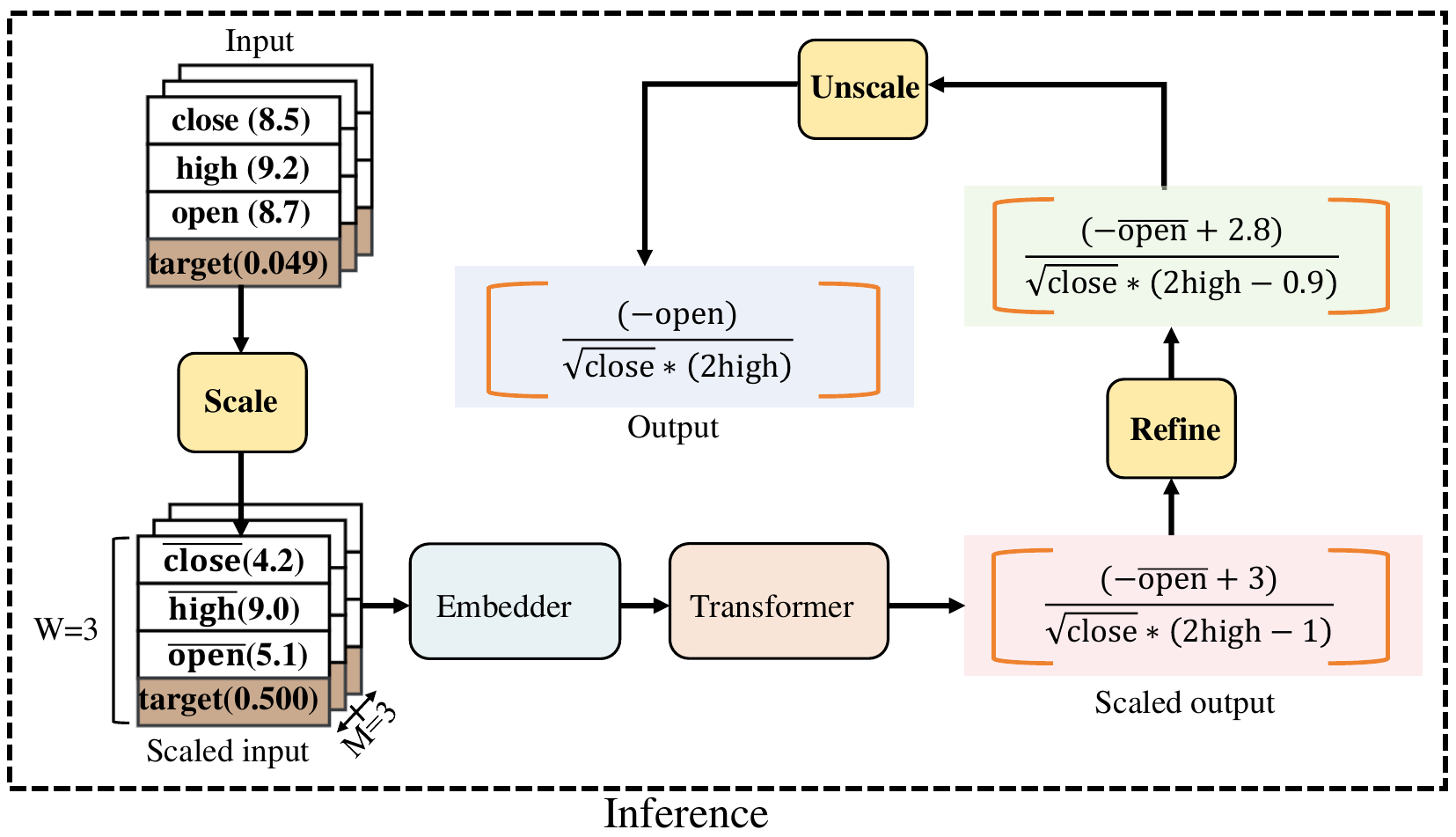}
  \end{minipage}
  \caption{Our risk factor-mining framework operates end-to-end. During training, the Embedder concatenates tokens from three sample pairs into a vector, which is then reduced via a two-layer, ReLU-activated Feed-Forward Network (FFN), producing M embedding vectors for the Transformer. In the inference stage, the complete HF risk factor expression is generated directly. We refine the constants in the factor formulas using the BFGS algorithm and incorporate a scaling process during inference to accommodate diverse HF feature sample points.}
  \label{fig:framework}
\end{figure*}

\section{The Method}
This section illustrates our IRFT for generating formulaic HF risk factors end-to-end. As shown in Fig. \ref{fig:framework}, our HF risk factor mining framework consists of two primary stages: Training and Inference.

\subsection{Model}
In this section, we introduce an Embedder to reduce input dimensions, making IRFT more suitable for high-dimensional transaction data. The Transformer model is sensitive to input samples when predicting factor expressions due to the complementarity of the attention heads. Some heads focus on extremes like $-1,0,1$ of exponential functions, while others concentrate on the periodicity of trigonometric functions.

\subsubsection{Embedder}
Our model is fed with $M$ HFT input points $(x,y) \in \mathbb{R}^{W+1}$, each of which is denoted as $3(W+1)$ tokens of dimension $d_{\text{emb}}$. As $W$ and $M$ increase, it results in long input sequences, e.g. 8400 tokens for $W=6$ and $M=400$, causing computational challenges ($O(n^2)$) for the Transformer. To alleviate this, we introduce an Encoder to map every input sample to a single embedding. Specifically, this embedder pads the empty input dimensions to $W_{\max}$, then feeds the $3(W_{\max}+1)*d_{\text{emb}}$-dimensional vector into a two-layer feed-forward network (FFN) with ReLU activations to reduce vector's dimension to $d_{\text{emb}}$. The $M$ embeddings of dimension $d_{\text{emb}}$ are then provided to the Transformer, effectively mitigating computational challenges.

\subsubsection{Transformer}
We utilize a sequence-to-sequence Transformer model, as proposed by  \cite{vaswani2017attention}, with 16 attention head and an embedding dimension of 512, totaling 84 million parameters. Following \cite{charton2021linear}, we observe that the solution to the factor mining task is an asymmetric model structure with a deeper decoder. In more details, we use 4 layers in the encoder and 16 layers in the decoder. The encoder effectively captures the distinctive features of HFT stock data, such as periodicity and continuity, by mixing short-ranged attention heads focusing on local features with long-ranged attention heads capturing the global form of the HF risk factor. Unlike previous risk factor mining models \cite{harman1976modern,lin2021deep}, our model accounts for the periodicity of stock sample points, capturing short-term volatility and cycling in the HFT stock market.

\subsection{Training.}
We employ the Adam optimizer to minimize the cross-entropy loss. Actually, our cross entropy loss calculates the probability distribution of target tokens and risk factor tokens, in order to find the factor token with the maximum probability of the target token. The learning rate is initially set to $2\times10^{-7}$ and is linearly increased to $2\times10^{-3}$ over the first 10,000 steps. Following the approach proposed by \cite{vaswani2017attention, xiang2025promptsculptor}, we then decay the learning rate as the inverse square root of the step numbers. To assess the model's performance, We take 10\% samples from the same generator as a validation set and trained IRFT model until the accuracy on the validation set was satisfied. We esitimate this process to span approximately 50 epochs, with each epoch consisting of 3 million sample points. Moreover, to minimize the padding waste, we put together sample point with similar lengths, which can ensure ensure more than 3M samples in each batch.

\subsection{Inference Tricks}
We describe four tricks to enhance IRFT’s inference performance. Previous SR models, like \cite{biggio2021neural,valipour2021symbolicgpt}, use a standard skeleton approach, predicting equation skeletons first and then fitting constants. These methods are less accurate and handle limited input dimensions ($W<3$). Our end-to-end (E2E) method predicts functions and constants simultaneously, allowing our model to process more input features ($W=7$) and handle large real-world datasets (about 26M samples).

\subsubsection{Refinement}
As suggested by \cite{kamienny2022end}, we significantly enhance effectiveness by adding a refinement step: fine-tuning constants using BFGS with our model predictions as initial values. This improves the skeleton method in two ways: providing stronger supervision signals by predicting complete HF risk factor expressions and enhancing constant accuracy with the BFGS algorithm, thereby boosting overall model performance.

\subsubsection{Scaling}
At the inference stage, we introduce a scaling procedure to ensure accurate predictive factor formulas from samples with varying means and variances. During training, we scale all HF input points to center their distributions at the origin with a variance of 1. For example, let $E$ be the inferred risk factor, $x$ be the input, $\mu = \text{mean}(x)$, and $\sigma = \text{std}(x)$. Then we replace $x$ with $\bar{x} = (x - \mu) / \sigma$, and the model predicts $\hat{E}(\bar{x}) = \hat{E}(\sigma x + \mu)$. This allows us to recover an approximate $E$ by unscaling the variables in $\hat{E}$. This approach makes IRFT insensitive to the scale of input points and represents constants outside $D_{(a,b)}$ as multiplications of constants within $D_{(a,b)}$. Unlike our method, DL-based approaches like \cite{petersen2019deep} often fail when samples fall outside the expected value range.

\subsubsection{Bagging and decoding}
With over 400 sample points, our experiment is less effective. To fully utilize extensive HFT data, we adopt bagging: if HFT input points $M$ exceed 400 during inference, we split the dataset into $B$ bags of 400 stock points each. We then generate $K$ formulaic factor candidates per bag, resulting in $BK$ candidates. Unlike previous language models, our approach generates multiple expressions for each bag.

\subsubsection{Inference time}
The speedup of our model inference comes from refining HF risk factor expression candidates. Given the large $BC$, we rank candidates by their error on input samples, remove duplicates or redundant factors, and keep the top $C$ candidates for refinement. To accelerate this process, we optimize using a subset of up to 1024 input samples. Parameters $B$, $K$, and $C$ can be fine-tuned for better speed and accuracy. In our experiment, we used $B=100$, $K=10$, $C=10$.

\begin{figure*}[ht]
  \centering
  \begin{minipage}{0.45\linewidth}   \includegraphics[width=\linewidth]{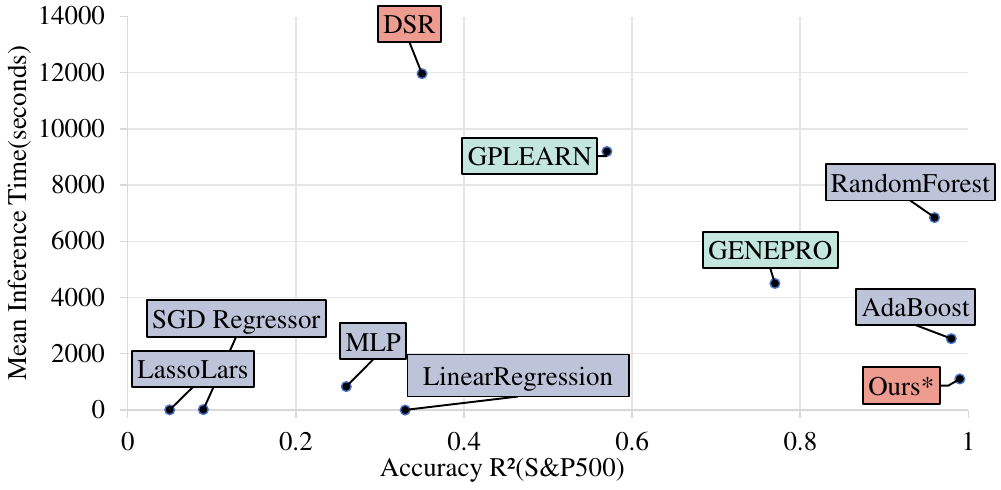}
  \end{minipage}
  \hfill
  \begin{minipage}{0.45\linewidth}
    \includegraphics[width=\linewidth]{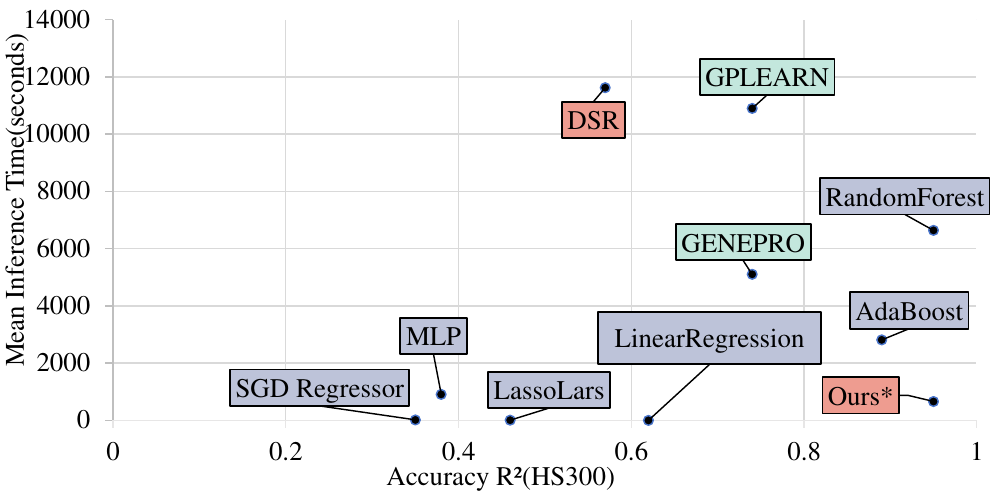}
  \end{minipage}
  \caption{The Pareto plot compares our framework with baselines like \textcolor{red}{DL-based methods}, \textcolor{blue}{ML-based methods}, and \textcolor{green}{GP-based methods} from SRBench about the average inference time and test performance.}
  \label{fig:pareto}
\end{figure*}

\begin{table*}[t]
    \centering
    \caption{Main results of S\&P500 Index and HS300 Index. Values after "$\pm$" represents the standard deviation, and the rest are the mean values, while the rest are mean values. “↑” indicates that higher values are better.}
    \resizebox{0.97\textwidth}{!}{%
    \begin{tabular}{ccccccc}
        \hline
        \multirow{2}{*}{Method} & \multicolumn{3}{c}{S\&P 500} & \multicolumn{3}{c}{HS 300} \\
        \cmidrule(r){2-4} \cmidrule(l){5-7}
         & $IC^*$ (↑) & Rank $IC^*$ (↑) & $IR^*$ (↑) & $IC^*$ (↑) & Rank $IC^*$ (↑) & $IR^*$ (↑)\\
        \midrule
        DSR & $0.0437\pm0.0054$ & $0.0453\pm0.0071$ & $0.2506\pm0.0246$ & $0.0427\pm0.0088$ & $0.0456\pm0.0059$ & $0.3395\pm0.0637$ \\
        GPLEARN & $0.0250\pm0.0099$ & $0.0547\pm0.0073$ & $0.4876\pm0.0307$ & $0.0494\pm0.0062$ & $0.0480\pm0.0063$ & $0.3600\pm0.0368$ \\
        GENEPRO & $0.0470\pm0.0068$ & $0.0550\pm0.0163$ & $0.2098\pm0.0339$ & $0.0444\pm0.0049$ & $0.0528\pm0.0090$ & $0.4787\pm0.0453$ \\
        Ours* & $\textbf{0.0662}\pm0.0077$ & $\textbf{0.0720}\pm0.0085$ & $\textbf{0.4960}\pm0.0817$ & $\textbf{0.0618}\pm0.0083$ & $\textbf{0.0683}\pm0.0088$ & $\textbf{0.6460}\pm0.1002$ \\
        \hline
    \end{tabular}%
    }
    \label{tab:IC_RankIC_IR}
\end{table*}

\section{EVALUATION}
In this section, we mainly answer the questions below: \textbf{Q1:} How does our proposed framework compare with other state-of-the-art methods based on SR tasks for generating HF risk factor expressions? \textbf{Q2:} How does our model perform as the size of HF risk factor collections increases? \textbf{Q3:} How does our framework perform in more realistic trading settings? 

\begin{table*}[t!]
\centering
\caption{Top 5 HF risk factor expressions based on $IC^*$ values in the HF risk factor collections (S\&P500 Index).}
\label{tab:factor}
\begin{tabular}{lcc}
\hline
No. & HF risk factor & $IC^*$ \\
\midrule
1 & $0.6*\log((2709.5-0.9\cdot\sin(0.4\cdot\text{close}-2.0)+16.7-40.1/((0.2\cdot\text{low}+0.4))))-0.4)+0.2$ & 0.0820 \\
2 & $-8.4+8.2/(-0.7*(-0.1-0.1/(1.9-0.5\cdot\text{close})))$ & 0.0641 \\
3 & $2.8*\sqrt{0.1/((-0.2\cdot\text{open}-11.7*(13.7-2.6\cdot\text{open})+1.0)^2)}-1.0$ & 0.0635 \\
4 & $1.3 * (0.2 * \lvert -0.1 \cdot \text{low} + 5.1 \rvert + 0.2) + 1.3$ & 0.0558 \\
5 & $0.8/((-0.1\cdot\text{open}-(0.1\cdot\text{high}+3.6)*(-3.0*(5.2-1.5\cdot\text{volume})*(40.0-7.0\cdot\text{low})+2046.9)))-0.5$ & 0.0522 \\
\hline
\end{tabular}
\end{table*}

% \begin{figure}[t]
%   \centering
%   \begin{minipage}{\columnwidth}
%     \centering
%     \includegraphics[width=1.2\linewidth]{AnonymousSubmission/LaTeX/box_S&P500.pdf}
%     \caption{Performance of different methods on S\&P500 Index (U.S. market).}
%   \end{minipage}
%   \begin{minipage}{\columnwidth}
%     \centering
%     \includegraphics[width=1.2\linewidth]{AnonymousSubmission/LaTeX/IC&RankIC&IR_S&P500.pdf}
%     \caption{A comparison of different factor generation methods using $IC^*$, $Rank IC^*$ and $IR^*$ for various factor pool sizes.}
%   \end{minipage}
%   \label{fig:box_sp500}
% \end{figure}

\begin{figure*}[t]
  \centering
  \begin{subfigure}[b]{0.7\textwidth}
    \centering
    \includegraphics[width=\linewidth]{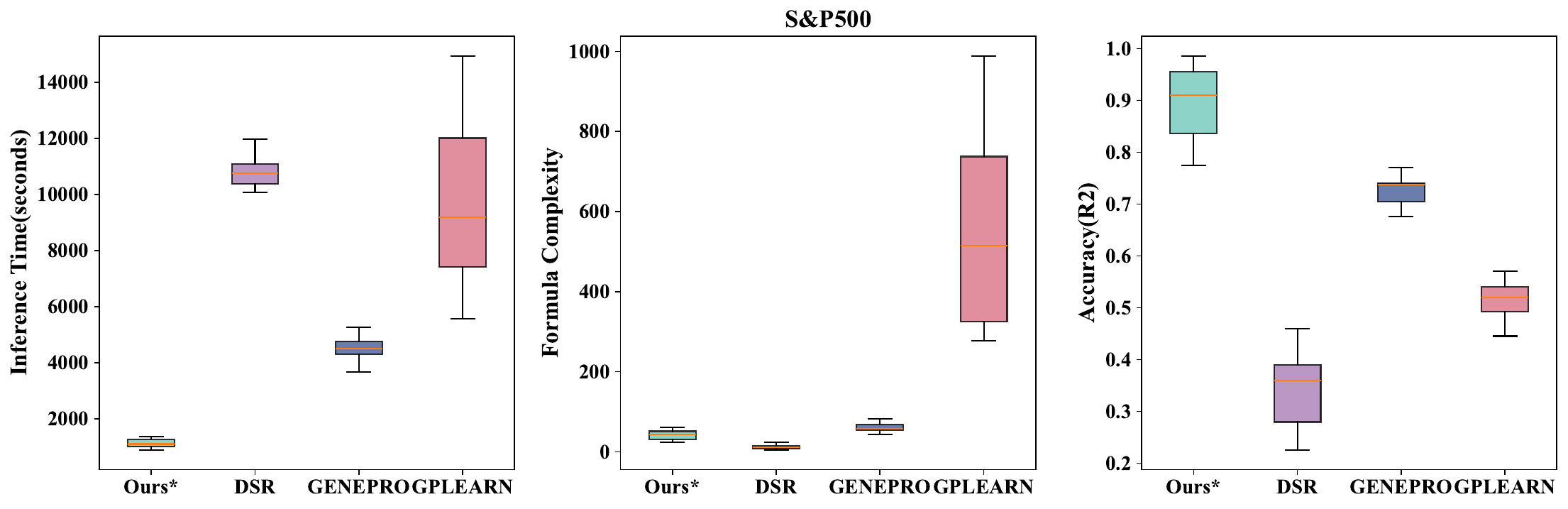}
  \end{subfigure}
  \hfill
  \begin{subfigure}[b]{0.7\textwidth}
    \centering
    \includegraphics[width=\linewidth]{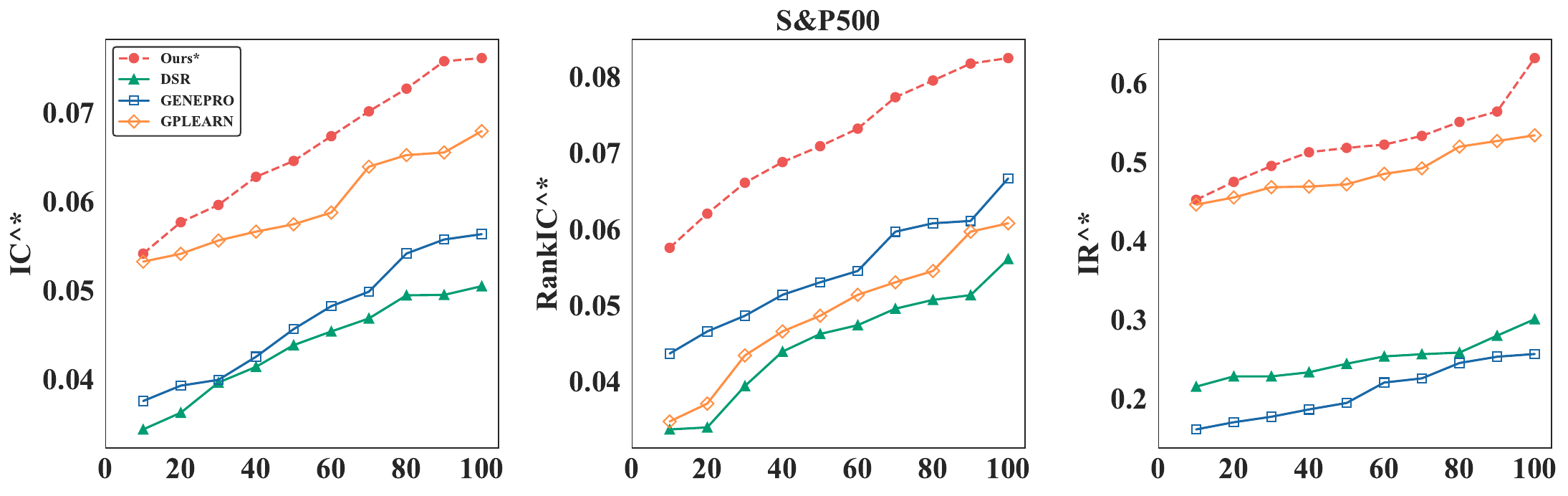}
  \end{subfigure}
  \caption{(a) Performance of different methods on S\&P500 Index (U.S. market). (b) A comparison of different factor generation methods using $IC^*$, $Rank IC^*$ and $IR^*$ for various factor pool sizes.}
  \label{fig:box_sp500}
\end{figure*}

\subsection{Experimental Settings}
\subsubsection{Datasets}
%\footnote{https://www.wind.com.cn/}
Our experiments are conducted on HFT data from the U.S. and China stock market (see Table \ref{tab:data}). The stock data was obtained from wind\footnote{https://www.wind.com.cn/}. Specifically, we analyze the $W$-dimensional trading data $X \in \mathbb{R}^W$ of
 the 1-min trading level of constituent stocks of the HS300 index and the S\&P500 index.
We preprocess raw stock  open/close/high/low/volume/vwap features using word expressions. It's important to highlight that our method employs an ascending variable sampling approach for features. This implies that instances such as $x_2 + x_3$ will not occur, but rather samples like $x_1 + x_2 + x_3$ will be generated. However, we have the flexibility to set the weight in front of $x_1$ to zero. The target value is the one-day-ahead RV, defined as $RV(t,j;n) = \sum_{j=1}^n (\ln P_{t,j} - \ln P_{t,j-1})^2$, where $n$ is the number of intraday trading intervals and $P_{i,j}$ is the closing price of the $j$-th interval of the $i$-th trading day \cite{andersen1998answering}. 

\begin{table}[t]
    \centering
    \caption{Market Data Overview}
    \resizebox{0.48\textwidth}{!}{%
    \begin{tabular}{lcc}
        \hline
         & The U.S. Market & The China Market \\
        \hline
        Training & 2023/01/03-2023/10/31 & 2022/10/31-2023/08/31 \\
        Evaluate & 2023/10/31-2023/12/29 & 2023/08/31-2023/10/31 \\
        Sample Size & 18,330,000 & 7,964,160 \\
        \hline
    \end{tabular}
    }
    \label{tab:data}
\end{table}

\subsubsection{Compared Approaches}
We compare IRFT with advanced methods from SRBench\cite{la2021contemporary} in generating HF risk factor expressions.
\begin{itemize}
  \item DL-based methods: \textbf{DSR}\cite{petersen2019deep} employs a parameterized neural network to model the distribution of a mathematical expression, represented as a sequence of operators and variables in the corresponding expression tree.
  \item GP-based method: \textbf{GPLEARN}\footnote{https://github.com/trevorstephens/gplearn} is a evolutionary algorithm specialized in SR, where the mathematical expressions of a population ‘evolve’ through the use of genetic operators such as mutation, crossover, and selection. \textbf{GENEPRO}\footnote{https://github.com/marcovirgolin/genepro} is another GP framework capable of handling not only regression feature variables but also environmental observations related to reinforcement learning (though we do not consider this option here). 
  \item ML-based methods: To evaluate the model more comprehensively, we compare IRFT with the end-to-end ML methods in SRBench. These methods take a week’s HF stock features as input and predict one-day-ahead RV. \textbf{AdaBoost} ensembles decision trees to directly predict stock trends. \textbf{Lasso Lars} is a linear regression method that combines Lasso and LARS. \textbf{Random Forest} is an ensemble learning method that averages the regression results from multiple decision trees. The other ML methods are \textbf{SGD Regressor} \cite{bottou2010large}, \textbf{MLP} and \textbf{Linear Regression}. These ML models do not generate formulaic HF risk factor expressions, but only output volatility prediction values. The hyperparameters are set by SRBench.
\end{itemize}

\subsubsection{Evaluation 
Metrics}
Each of the following four positive indicators assesses the correlation between actual stock price volatility and future volatility forecasts.
\paragraph{$\boldsymbol{R^2}$}
It measures the explanatory degree of one-day-ahead RV variation for factor-mining tasks.

\begin{equation}
R^2 = 1 - \frac{\sum_{i=1}^{M_{\text{test}}} (y_i - \hat{y}_i)^2}{\sum_{i=1}^{M_{\text{test}}} (y_i - \bar{y})^2} \label{eq:1}
\end{equation}
\paragraph{$\boldsymbol{IC^*}$}
It is the cross-section correlation coefficient of the HF risk factor values and the one-day-ahead RV values of trading stocks, which ranges from $[-1,1]$. 
\begin{equation}
IC^* = \bar{\sigma} \left( \sum_{i=1}^k w_i f_i(X), y_i \right) \label{eq:2}
\end{equation}
Where, $\sigma(f_l(X), f_k(X)) = \frac{\sum_{l=1}^n (f_{tl} - \bar{f}_t)(f_{tk} - \bar{f}_t)}{\sqrt{\sum_{l=1}^n (f_{tl} - \bar{f}_t)^2 \sum_{k=1}^n (f_{tk} - \bar{f}_t)^2}}$, $y_i$ is the one-day-ahead RV. We simplify the calculation by taking the daily average of the Pearson correlation coefficient between the HF risk factors and the targets over all trading days. 
$\bar{\sigma} \left( \sum_{i=1}^k w_i f_i(X), y_i \right) = E_t \left[ \sigma \left( \sum_{i=1}^k w_i f_i(X), y_i \right) \right]$ and $\bar{\sigma} \left( \sum_{i=1}^k w_i f_i(X), y_i \right) \in [-1,1]$.
\paragraph{$\boldsymbol{Rank IC^*}$} It is $IC^*$ ranking of the HF risk factor values and the one-day-ahead RV of the selected stocks.
\begin{equation}
Rank IC^* = \bar{\sigma} \left( r \left( \sum_{i=1}^k w_i f_i(X) \right), r(y_i) \right) \label{eq:3}
\end{equation}
Where, $r(\cdot)$ is the ranking operator of the HF risk factor value and the one-day-ahead RV.
\paragraph{$\boldsymbol{IR^*}$}
It is the mean of the $IC^*$ divided by the standard deviation of these $IC^*$.
\begin{equation}
IR^* = \frac{\bar{Ret}_t^{IC}}{\text{std}(Ret_t^{IC})} \label{eq:4}
\end{equation}

% \begin{figure*}[h]
%     \centering   \includegraphics[width=0.7\linewidth]{box_S_P500.pdf}
%     \caption{Performance of different methods on  S\_P500 Index (U.S. market).}
%     \label{box_sp500}
% \end{figure*}

\begin{figure*}[t]
  \centering
  \begin{minipage}{0.35\textwidth} 
    \includegraphics[width=\linewidth]{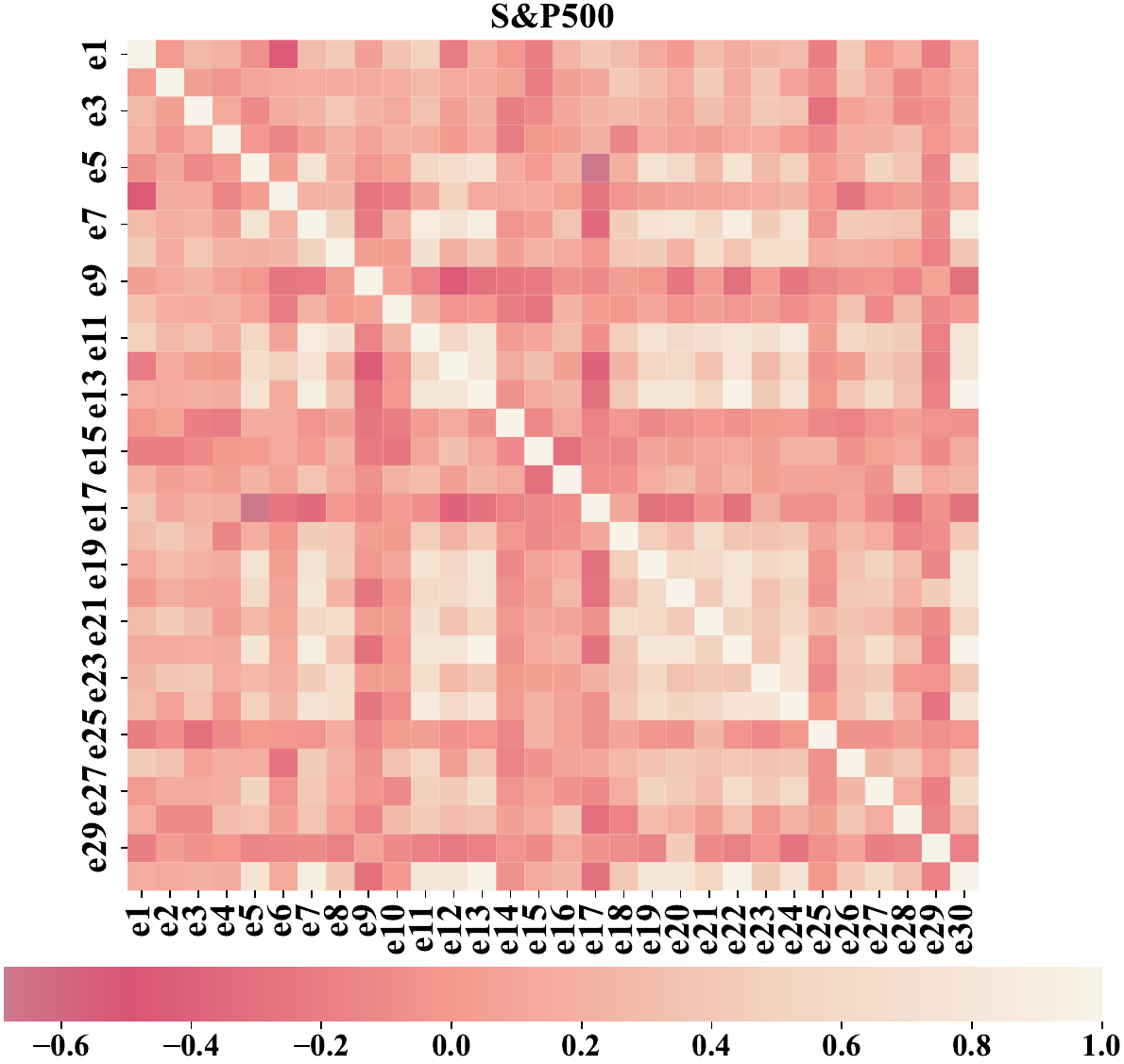}
  \end{minipage}
  \begin{minipage}{0.325\textwidth} 
    \includegraphics[width=\linewidth]{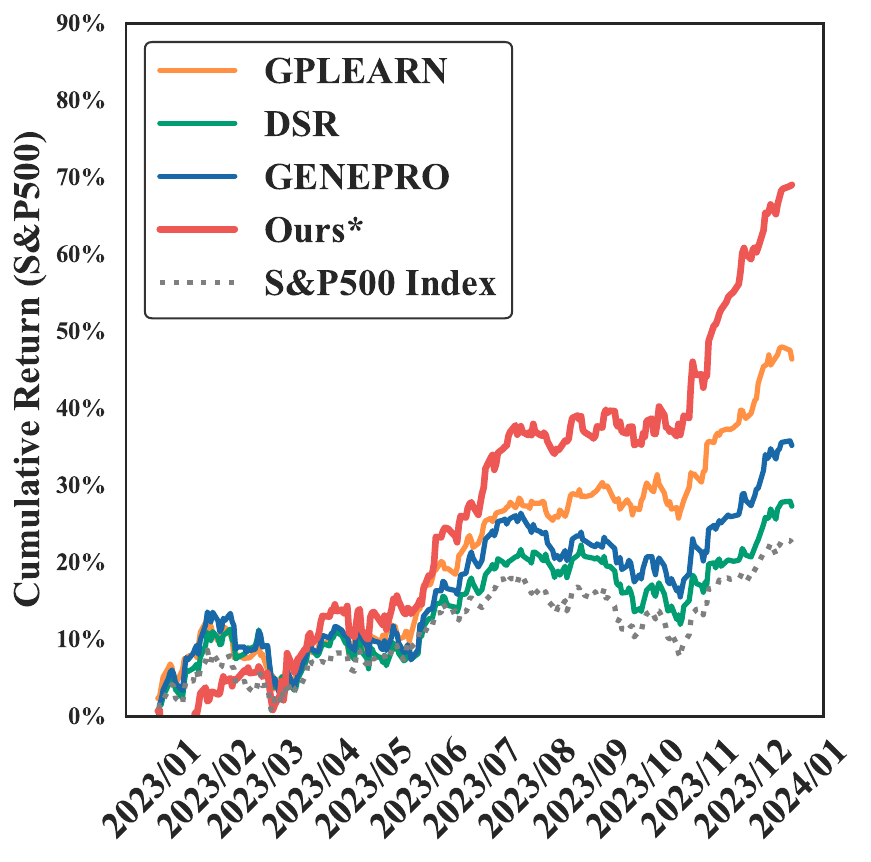}
  \end{minipage}
  \caption{(a) The correlation of the HF risk factor collections on S\&P500 Index. (b) Portfolio simulations on the S\&P500 Index: a backtesting comparison.}
  \label{fig:heatmap_sp500}
\end{figure*}

\subsection{Comparison across all HF risk factor generators (RQ1)}
We evaluate the performance of three baseline methods on HFT data sets derived from the constituents of the S\&P500 index. The results, in terms of inference time, expression complexity, and prediction performance ($R^2$), are presented using box plots in Fig. \ref{fig:box_sp500}(a).

Our method demonstrates a significant advantage in inference time, being 17 times faster than DSR, and exhibits a narrow distribution, indicating high stability and efficiency. This is attributed to our method's use of the efficient encoding and decoding mechanism, which transforms the features and target variables in the data sets into the input and output of HF risk factor expression tokens, thereby reducing the search space and computation of the symbolic solver. GPLEARN, while faster than DSR, has a more dispersed distribution, suggesting its performance is more sensitive to the data sets. This could be due to GPLEARN's use of genetic programming to search for symbolic expressions, which often results in overly long and complex expressions and lacks effective regularization mechanisms. In terms of expression complexity, DSR generates the most concise expressions, possibly due to its use of a regularization term to penalize excessively long and complex expressions, thus avoiding overfitting and redundancy. Our method and DSR have similar expression complexity, which is significantly lower than GPLEARN. GPLEARN generates the longest expressions, with an average length of 512 and a maximum length of 989. Regarding prediction performance, our method generates HF risk factor expressions with an $R^2$ value of 0.92, significantly outperforming other methods. This suggests that our method can accurately fit and predict the volatility trend of the stock data sets. GPLEARN and GENEPRO are the second best methods, with comparable $R^2$ values. Moreover, DSR performs the worst, with a low and large $R^2$ variation, ranging from 0.23 to 0.46. Moreover, Fig. \ref{fig:pareto} demonstrates that our framework surpasses other 10 baselines in SRBench with the highest performance and shortest inference time. .

Table \ref{tab:IC_RankIC_IR} shows that, compared to other baseline methods, our method generates HF risk factors with the highest values of $IC^*$, $Rank IC^*$, and $IR^*$, which assess the predictive ability of the generated HF risk factor expressions, like their ability to capture the future stock volatility trend. For the indicators $IC^*$ and $Rank IC^*$ , DSR and GENEPRO perform similarly on both the S\&P500 and HS300 index data sets. For the indicator $IR^*$, the GP methods based on SR tasks significantly outperform  DSR.

\subsection{Comparison of formulaic generators with varying pool capacity (RQ2)}
Fig. \ref{fig:box_sp500} presents the mean values of $IC^*$, $Rank IC^*$, and $IR^*$ for the HF risk factor collections generated by different methods under varying factor pool sizes. From this figure, our framework, IRFT, consistently exhibits the highest $IC^*$, $Rank IC^*$ and $IR^*$ values across all factor pool sizes. Furthermore, IRFT demonstrates scalability with respect to the factor pool size. That is, as the factor pool size increases, it continues to identify superior HF risk factor expressions without succumbing to overfitting or redundancy issues. This suggests that IRFT can effectively leverage the semantic information of the large language model, as well as the flexibility of symbolic regression, to generate diverse and innovative HF risk factors.

\subsection{Investment Simulation (RQ3)}
During backtesting period, we analyze factor set heat maps (Fig. \ref{fig:heatmap_sp500}(a) and select 10 independent risk factors(the more information they contain). We calculate risk factor scores for all assets in the portfolio ($k=30$ stocks in our paper) for each adjustment period (each trading day). The factor score for the $s$th stock is $a(s) = \sum_{n=1}^N w(n) V(s,n)$, where $w(n)$ is the weight of the $n$th factor ($\text{IC}^*$ value) and $V(s,n)$ is the $n$th factor value of the $s$th stock. These factors help select high-yielding stocks for the portfolio. As shown in Fig. \ref{fig:heatmap_sp500}(b), we record cumulative net returns over a one-year test period. Despite the fact that our framework does not explicitly optimize for absolute return, it still performs well in the backtest, even during the test period when the international environment is unstable and the domestic economy slows down. Compared with other methods, our framework can still achieve the highest profit, exceeding the second-place method GENEPRO by 21\%. Table \ref{tab:factor} presents the top five HF risk factors selected from the HF risk factor collections, based on their $IC^*$ values. These factors, which include arbitrary constant terms, are highly flexible and can be adapted to any range of values. For simplicity, we round the constant terms to one decimal place. 

\section{Related Work}
\noindent\textbf{Symbolic Regression.} Symbolic Regression (SR) focuses on identifying mathematical expressions that best describe relationships \cite{kamienny2022end}. Mining quantitative factors is essentially an SR task, extracting accurate mathematical patterns from historical trading data. The dominant approach is Genetic Programming (GP) \cite{cui2021alphaevolve}, but it has two drawbacks: slow execution \cite{langdon2011graphics}, due to the expansive search space requiring numerous generations to converge, and a tendency towards overfitting \cite{fitzgerald2013bootstrapping}, dependent on input fitness. Given the time-sensitive nature of quantitative trading, we aim to generate risk factors that effectively capture short-term market volatility.

\noindent\textbf{Language Models.} 
The pre-trained Finbert\cite{liu2021finbert} or other pre-trained Fingpt\cite{yang2023fingpt,liu2023fingpt,xu2025learning, xu2025finmultitime} mainly focus on semantic processing. However, the task of generating formulaic factors primarily focuses on tackling numbers and symbols, rather than comprehending extensive vocabularies. They encounter difficulties in complex reasoning tasks, like predicting formulas for large numbers\cite{mishra2022numglue}. In pretraining, models struggle with rarely or never seen symbols. And they can only deal with sequences. We treat HF risk factors as a language for explaining market volatility trends and train a Transformer. This model offers two benefits: They are fast because of leveraging previous experience with inference being a single forward pass. And they are less over fitting-prone, as they do not require loss optimization based on inputs.

\section{Conclusion}
In this paper, we introduce a transformer model for HF risk factor mining, replacing traditional manual factor construction. Unlike previous language models that focus on word relationships for natural language comprehension, our model uses a hybrid symbolic-numeric vocabulary. Symbolic tokens represent operators/stock features, while numeric tokens denote constants. The Transformer predicts complete formulaic factors end-to-end and refines constants using a non-convex optimizer. Our framework outperforms 10 approaches in SRBench and scales to a 6-dimensional, 26 million HF dataset. Extensive experiments show that IRFT surpasses formulaic risk factor mining baselines, achieving a 30\% excess investment return on HS300 and S\&P500 datasets and significantly speeding up inference times.

\begin{acks}
This research was supported by the Beijing Natural Science Foundation (Grant No. 4232039), the Youth Fund of the Computer Network Information Center, Chinese Academy of Sciences (Grant No. 24YF07), and the Strategic Priority Research Program of the Chinese Academy of Sciences (Grant No. XDB0500103).
\end{acks}

\bibliographystyle{ACM-Reference-Format}
\bibliography{sample-base}


\begin{thebibliography}{30}

%%% the Web address.

\ifx \showCODEN    \undefined \def \showCODEN     #1{\unskip}     \fi
\ifx \showISBNx    \undefined \def \showISBNx     #1{\unskip}     \fi
\ifx \showISBNxiii \undefined \def \showISBNxiii  #1{\unskip}     \fi
\ifx \showISSN     \undefined \def \showISSN      #1{\unskip}     \fi
\ifx \showLCCN     \undefined \def \showLCCN      #1{\unskip}     \fi
\ifx \shownote     \undefined \def \shownote      #1{#1}          \fi
\ifx \showarticletitle \undefined \def \showarticletitle #1{#1}   \fi
\ifx \showURL      \undefined \def \showURL       {\relax}        \fi
% invisible to TeX
\providecommand\bibfield[2]{#2}
\providecommand\bibinfo[2]{#2}
\providecommand\natexlab[1]{#1}
\providecommand\showeprint[2][]{arXiv:#2}

\bibitem[Andersen and Bollerslev(1998)]%
        {andersen1998answering}
\bibfield{author}{\bibinfo{person}{Torben~G Andersen} {and} \bibinfo{person}{Tim Bollerslev}.} \bibinfo{year}{1998}\natexlab{}.
\newblock \showarticletitle{Answering the skeptics: Yes, standard volatility models do provide accurate forecasts}.
\newblock \bibinfo{journal}{\emph{International economic review}} (\bibinfo{year}{1998}), \bibinfo{pages}{885--905}.
\newblock


\bibitem[Ang et~al\mbox{.}(2020)]%
        {ang2020using}
\bibfield{author}{\bibinfo{person}{Andrew Ang}, \bibinfo{person}{Jun Liu}, {and} \bibinfo{person}{Krista Schwarz}.} \bibinfo{year}{2020}\natexlab{}.
\newblock \showarticletitle{Using stocks or portfolios in tests of factor models}.
\newblock \bibinfo{journal}{\emph{Journal of Financial and Quantitative Analysis}} \bibinfo{volume}{55}, \bibinfo{number}{3} (\bibinfo{year}{2020}), \bibinfo{pages}{709--750}.
\newblock


\bibitem[Becker et~al\mbox{.}(2023)]%
        {becker2023predicting}
\bibfield{author}{\bibinfo{person}{S{\"o}ren Becker}, \bibinfo{person}{Michal Klein}, \bibinfo{person}{Alexander Neitz}, \bibinfo{person}{Giambattista Parascandolo}, {and} \bibinfo{person}{Niki Kilbertus}.} \bibinfo{year}{2023}\natexlab{}.
\newblock \showarticletitle{Predicting ordinary differential equations with transformers}. In \bibinfo{booktitle}{\emph{International Conference on Machine Learning}}. PMLR, \bibinfo{pages}{1978--2002}.
\newblock


\bibitem[Biggio et~al\mbox{.}(2021)]%
        {biggio2021neural}
\bibfield{author}{\bibinfo{person}{Luca Biggio}, \bibinfo{person}{Tommaso Bendinelli}, \bibinfo{person}{Alexander Neitz}, \bibinfo{person}{Aurelien Lucchi}, {and} \bibinfo{person}{Giambattista Parascandolo}.} \bibinfo{year}{2021}\natexlab{}.
\newblock \showarticletitle{Neural symbolic regression that scales}. In \bibinfo{booktitle}{\emph{International Conference on Machine Learning}}. PMLR, \bibinfo{pages}{936--945}.
\newblock


\bibitem[Bottou(2010)]%
        {bottou2010large}
\bibfield{author}{\bibinfo{person}{L{\'e}on Bottou}.} \bibinfo{year}{2010}\natexlab{}.
\newblock \showarticletitle{Large-scale machine learning with stochastic gradient descent}. In \bibinfo{booktitle}{\emph{Proceedings of COMPSTAT'2010: 19th International Conference on Computational StatisticsParis France, August 22-27, 2010 Keynote, Invited and Contributed Papers}}. Springer, \bibinfo{pages}{177--186}.
\newblock


\bibitem[Carhart(1997)]%
        {carhart1997persistence}
\bibfield{author}{\bibinfo{person}{Mark~M Carhart}.} \bibinfo{year}{1997}\natexlab{}.
\newblock \showarticletitle{On persistence in mutual fund performance}.
\newblock \bibinfo{journal}{\emph{The Journal of finance}} \bibinfo{volume}{52}, \bibinfo{number}{1} (\bibinfo{year}{1997}), \bibinfo{pages}{57--82}.
\newblock


\bibitem[Charton(2021)]%
        {charton2021linear}
\bibfield{author}{\bibinfo{person}{Fran{\c{c}}ois Charton}.} \bibinfo{year}{2021}\natexlab{}.
\newblock \showarticletitle{Linear algebra with transformers}.
\newblock \bibinfo{journal}{\emph{arXiv preprint arXiv:2112.01898}} (\bibinfo{year}{2021}).
\newblock


\bibitem[Cui et~al\mbox{.}(2021)]%
        {cui2021alphaevolve}
\bibfield{author}{\bibinfo{person}{Can Cui}, \bibinfo{person}{Wei Wang}, \bibinfo{person}{Meihui Zhang}, \bibinfo{person}{Gang Chen}, \bibinfo{person}{Zhaojing Luo}, {and} \bibinfo{person}{Beng~Chin Ooi}.} \bibinfo{year}{2021}\natexlab{}.
\newblock \showarticletitle{AlphaEvolve: A Learning Framework to Discover Novel Alphas in Quantitative Investment}. In \bibinfo{booktitle}{\emph{Proceedings of the 2021 International Conference on Management of Data}}. \bibinfo{pages}{2208--2216}.
\newblock


\bibitem[Fama and French(1992)]%
        {fama1992cross}
\bibfield{author}{\bibinfo{person}{Eugene~F Fama} {and} \bibinfo{person}{Kenneth~R French}.} \bibinfo{year}{1992}\natexlab{}.
\newblock \showarticletitle{The cross-section of expected stock returns}.
\newblock \bibinfo{journal}{\emph{the Journal of Finance}} \bibinfo{volume}{47}, \bibinfo{number}{2} (\bibinfo{year}{1992}), \bibinfo{pages}{427--465}.
\newblock


\bibitem[Fan et~al\mbox{.}(2021)]%
        {fan2021augmented}
\bibfield{author}{\bibinfo{person}{Jianqing Fan}, \bibinfo{person}{Yuan Ke}, {and} \bibinfo{person}{Yuan Liao}.} \bibinfo{year}{2021}\natexlab{}.
\newblock \showarticletitle{Augmented factor models with applications to validating market risk factors and forecasting bond risk premia}.
\newblock \bibinfo{journal}{\emph{Journal of Econometrics}} \bibinfo{volume}{222}, \bibinfo{number}{1} (\bibinfo{year}{2021}), \bibinfo{pages}{269--294}.
\newblock


\bibitem[Fitzgerald et~al\mbox{.}(2013)]%
        {fitzgerald2013bootstrapping}
\bibfield{author}{\bibinfo{person}{Jeannie Fitzgerald}, \bibinfo{person}{R~Muhammad~Atif Azad}, {and} \bibinfo{person}{Conor Ryan}.} \bibinfo{year}{2013}\natexlab{}.
\newblock \showarticletitle{A bootstrapping approach to reduce over-fitting in genetic programming}. In \bibinfo{booktitle}{\emph{Proceedings of the 15th annual conference companion on Genetic and evolutionary computation}}. \bibinfo{pages}{1113--1120}.
\newblock


\bibitem[Harman(1976)]%
        {harman1976modern}
\bibfield{author}{\bibinfo{person}{Harry~Horace Harman}.} \bibinfo{year}{1976}\natexlab{}.
\newblock \bibinfo{booktitle}{\emph{Modern factor analysis}}.
\newblock \bibinfo{publisher}{University of Chicago press}.
\newblock


\bibitem[Kamienny et~al\mbox{.}(2022)]%
        {kamienny2022end}
\bibfield{author}{\bibinfo{person}{Pierre-Alexandre Kamienny}, \bibinfo{person}{St{\'e}phane d'Ascoli}, \bibinfo{person}{Guillaume Lample}, {and} \bibinfo{person}{Fran{\c{c}}ois Charton}.} \bibinfo{year}{2022}\natexlab{}.
\newblock \showarticletitle{End-to-end symbolic regression with transformers}.
\newblock \bibinfo{journal}{\emph{Advances in Neural Information Processing Systems}}  \bibinfo{volume}{35} (\bibinfo{year}{2022}), \bibinfo{pages}{10269--10281}.
\newblock


\bibitem[La~Cava et~al\mbox{.}(2021)]%
        {la2021contemporary}
\bibfield{author}{\bibinfo{person}{William La~Cava}, \bibinfo{person}{Patryk Orzechowski}, \bibinfo{person}{Bogdan Burlacu}, \bibinfo{person}{Fabr{\'\i}cio~Olivetti de Fran{\c{c}}a}, \bibinfo{person}{Marco Virgolin}, \bibinfo{person}{Ying Jin}, \bibinfo{person}{Michael Kommenda}, {and} \bibinfo{person}{Jason~H Moore}.} \bibinfo{year}{2021}\natexlab{}.
\newblock \showarticletitle{Contemporary symbolic regression methods and their relative performance}.
\newblock \bibinfo{journal}{\emph{arXiv preprint arXiv:2107.14351}} (\bibinfo{year}{2021}).
\newblock


\bibitem[Lample and Charton(2019)]%
        {lample2019deep}
\bibfield{author}{\bibinfo{person}{Guillaume Lample} {and} \bibinfo{person}{Fran{\c{c}}ois Charton}.} \bibinfo{year}{2019}\natexlab{}.
\newblock \showarticletitle{Deep learning for symbolic mathematics}.
\newblock \bibinfo{journal}{\emph{arXiv preprint arXiv:1912.01412}} (\bibinfo{year}{2019}).
\newblock


\bibitem[Langdon(2011)]%
        {langdon2011graphics}
\bibfield{author}{\bibinfo{person}{William~B Langdon}.} \bibinfo{year}{2011}\natexlab{}.
\newblock \showarticletitle{Graphics processing units and genetic programming: an overview}.
\newblock \bibinfo{journal}{\emph{Soft computing}}  \bibinfo{volume}{15} (\bibinfo{year}{2011}), \bibinfo{pages}{1657--1669}.
\newblock


\bibitem[Lin et~al\mbox{.}(2021)]%
        {lin2021deep}
\bibfield{author}{\bibinfo{person}{Hengxu Lin}, \bibinfo{person}{Dong Zhou}, \bibinfo{person}{Weiqing Liu}, {and} \bibinfo{person}{Jiang Bian}.} \bibinfo{year}{2021}\natexlab{}.
\newblock \showarticletitle{Deep risk model: a deep learning solution for mining latent risk factors to improve covariance matrix estimation}. In \bibinfo{booktitle}{\emph{Proceedings of the Second ACM International Conference on AI in Finance}}. \bibinfo{pages}{1--8}.
\newblock


\bibitem[Liu et~al\mbox{.}(2023)]%
        {liu2023fingpt}
\bibfield{author}{\bibinfo{person}{Xiao-Yang Liu}, \bibinfo{person}{Guoxuan Wang}, {and} \bibinfo{person}{Daochen Zha}.} \bibinfo{year}{2023}\natexlab{}.
\newblock \showarticletitle{Fingpt: Democratizing internet-scale data for financial large language models}.
\newblock \bibinfo{journal}{\emph{@article{zhang2023instruct, title={Instruct-fingpt: Financial sentiment analysis by instruction tuning of general-purpose large language models}, author={Zhang, Boyu and Yang, Hongyang and Liu, Xiao-Yang}, journal={arXiv preprint arXiv:2306.12659}, year={2023} }arXiv preprint arXiv:2307.10485}} (\bibinfo{year}{2023}).
\newblock


\bibitem[Liu et~al\mbox{.}(2021)]%
        {liu2021finbert}
\bibfield{author}{\bibinfo{person}{Zhuang Liu}, \bibinfo{person}{Degen Huang}, \bibinfo{person}{Kaiyu Huang}, \bibinfo{person}{Zhuang Li}, {and} \bibinfo{person}{Jun Zhao}.} \bibinfo{year}{2021}\natexlab{}.
\newblock \showarticletitle{Finbert: A pre-trained financial language representation model for financial text mining}. In \bibinfo{booktitle}{\emph{Proceedings of the twenty-ninth international conference on international joint conferences on artificial intelligence}}. \bibinfo{pages}{4513--4519}.
\newblock


\bibitem[Mishra et~al\mbox{.}(2022)]%
        {mishra2022numglue}
\bibfield{author}{\bibinfo{person}{Swaroop Mishra}, \bibinfo{person}{Arindam Mitra}, \bibinfo{person}{Neeraj Varshney}, \bibinfo{person}{Bhavdeep Sachdeva}, \bibinfo{person}{Peter Clark}, \bibinfo{person}{Chitta Baral}, {and} \bibinfo{person}{Ashwin Kalyan}.} \bibinfo{year}{2022}\natexlab{}.
\newblock \showarticletitle{NumGLUE: A suite of fundamental yet challenging mathematical reasoning tasks}.
\newblock \bibinfo{journal}{\emph{arXiv preprint arXiv:2204.05660}} (\bibinfo{year}{2022}).
\newblock


\bibitem[Petersen et~al\mbox{.}(2019)]%
        {petersen2019deep}
\bibfield{author}{\bibinfo{person}{Brenden~K Petersen}, \bibinfo{person}{Mikel Landajuela}, \bibinfo{person}{T~Nathan Mundhenk}, \bibinfo{person}{Claudio~P Santiago}, \bibinfo{person}{Soo~K Kim}, {and} \bibinfo{person}{Joanne~T Kim}.} \bibinfo{year}{2019}\natexlab{}.
\newblock \showarticletitle{Deep symbolic regression: Recovering mathematical expressions from data via risk-seeking policy gradients}.
\newblock \bibinfo{journal}{\emph{arXiv preprint arXiv:1912.04871}} (\bibinfo{year}{2019}).
\newblock


\bibitem[Prices(1964)]%
        {prices1964theory}
\bibfield{author}{\bibinfo{person}{Capital~Asset Prices}.} \bibinfo{year}{1964}\natexlab{}.
\newblock \showarticletitle{A theory of Market Equilibrium under Conditions of Risk}.
\newblock \bibinfo{journal}{\emph{Journal of Finance}} \bibinfo{volume}{19}, \bibinfo{number}{3} (\bibinfo{year}{1964}), \bibinfo{pages}{425--444}.
\newblock


\bibitem[Sheikh(1996)]%
        {sheikh1996barra}
\bibfield{author}{\bibinfo{person}{Aamir Sheikh}.} \bibinfo{year}{1996}\natexlab{}.
\newblock \showarticletitle{BARRA’s risk models}.
\newblock \bibinfo{journal}{\emph{Barra Research Insights}} (\bibinfo{year}{1996}), \bibinfo{pages}{1--24}.
\newblock


\bibitem[Valipour et~al\mbox{.}(2021)]%
        {valipour2021symbolicgpt}
\bibfield{author}{\bibinfo{person}{Mojtaba Valipour}, \bibinfo{person}{Bowen You}, \bibinfo{person}{Maysum Panju}, {and} \bibinfo{person}{Ali Ghodsi}.} \bibinfo{year}{2021}\natexlab{}.
\newblock \showarticletitle{Symbolicgpt: A generative transformer model for symbolic regression}.
\newblock \bibinfo{journal}{\emph{arXiv preprint arXiv:2106.14131}} (\bibinfo{year}{2021}).
\newblock


\bibitem[Vaswani et~al\mbox{.}(2017)]%
        {vaswani2017attention}
\bibfield{author}{\bibinfo{person}{Ashish Vaswani}, \bibinfo{person}{Noam Shazeer}, \bibinfo{person}{Niki Parmar}, \bibinfo{person}{Jakob Uszkoreit}, \bibinfo{person}{Llion Jones}, \bibinfo{person}{Aidan~N Gomez}, \bibinfo{person}{{\L}ukasz Kaiser}, {and} \bibinfo{person}{Illia Polosukhin}.} \bibinfo{year}{2017}\natexlab{}.
\newblock \showarticletitle{Attention is all you need}.
\newblock \bibinfo{journal}{\emph{Advances in neural information processing systems}}  \bibinfo{volume}{30} (\bibinfo{year}{2017}).
\newblock


\bibitem[Xiang et~al\mbox{.}(2025)]%
        {xiang2025promptsculptor}
\bibfield{author}{\bibinfo{person}{Dawei Xiang}, \bibinfo{person}{Wenyan Xu}, \bibinfo{person}{Kexin Chu}, \bibinfo{person}{Zixu Shen}, \bibinfo{person}{Tianqi Ding}, {and} \bibinfo{person}{Wei Zhang}.} \bibinfo{year}{2025}\natexlab{}.
\newblock \showarticletitle{PromptSculptor: Multi-Agent Based Text-to-Image Prompt Optimization}.
\newblock \bibinfo{journal}{\emph{arXiv preprint arXiv:2509.12446}} (\bibinfo{year}{2025}).
\newblock


\bibitem[Xu et~al\mbox{.}(2025a)]%
        {xu2025mining}
\bibfield{author}{\bibinfo{person}{Wenyan Xu}, \bibinfo{person}{Jiayu Chen}, \bibinfo{person}{Dawei Xiang}, \bibinfo{person}{Chen Li}, \bibinfo{person}{Yonghong Hu}, {and} \bibinfo{person}{Zhonghua Lu}.} \bibinfo{year}{2025}\natexlab{a}.
\newblock \showarticletitle{Mining Intraday Risk Factor Collections via Hierarchical Reinforcement Learning based on Transferred Options}.
\newblock \bibinfo{journal}{\emph{arXiv preprint arXiv:2501.07274}} (\bibinfo{year}{2025}).
\newblock


\bibitem[Xu et~al\mbox{.}(2025b)]%
        {xu2025finmultitime}
\bibfield{author}{\bibinfo{person}{Wenyan Xu}, \bibinfo{person}{Dawei Xiang}, \bibinfo{person}{Yue Liu}, \bibinfo{person}{Xiyu Wang}, \bibinfo{person}{Yanxiang Ma}, \bibinfo{person}{Liang Zhang}, \bibinfo{person}{Chang Xu}, {and} \bibinfo{person}{Jiaheng Zhang}.} \bibinfo{year}{2025}\natexlab{b}.
\newblock \showarticletitle{FinMultiTime: A Four-Modal Bilingual Dataset for Financial Time-Series Analysis}.
\newblock \bibinfo{journal}{\emph{arXiv preprint arXiv:2506.05019}} (\bibinfo{year}{2025}).
\newblock


\bibitem[Xu et~al\mbox{.}(2025c)]%
        {xu2025learning}
\bibfield{author}{\bibinfo{person}{Wenyan Xu}, \bibinfo{person}{Dawei Xiang}, \bibinfo{person}{Rundong Wang}, \bibinfo{person}{Yonghong Hu}, \bibinfo{person}{Liang Zhang}, \bibinfo{person}{Jiayu Chen}, {and} \bibinfo{person}{Zhonghua Lu}.} \bibinfo{year}{2025}\natexlab{c}.
\newblock \showarticletitle{Learning Explainable Stock Predictions with Tweets Using Mixture of Experts}.
\newblock \bibinfo{journal}{\emph{arXiv preprint arXiv:2507.20535}} (\bibinfo{year}{2025}).
\newblock


\bibitem[Yang et~al\mbox{.}(2023)]%
        {yang2023fingpt}
\bibfield{author}{\bibinfo{person}{Hongyang Yang}, \bibinfo{person}{Xiao-Yang Liu}, {and} \bibinfo{person}{Christina~Dan Wang}.} \bibinfo{year}{2023}\natexlab{}.
\newblock \showarticletitle{Fingpt: Open-source financial large language models}.
\newblock \bibinfo{journal}{\emph{arXiv preprint arXiv:2306.06031}} (\bibinfo{year}{2023}).
\newblock


\end{thebibliography}

\appendix

\section{APPENDIX}

\begin{table}[t]
\centering
\caption{List of all operators employed within our framework.}
\label{tab:operators}
\begin{tabular}{ccc}
\toprule
Category & Operator & Description \\
\midrule
\multirow{4}{*}{\centering CS-B} & add(x,y) & \multirow{4}{*}{\centering Arithmetic operators} \\
& sub(x,y) & \\
& mul(x,y) & \\
& div(x,y) & \\
\multirow{9}{*}{\centering CS-U} & inv(x) & The inverse of x \\
& sqr(x) & The square of x \\
& sqrt(x) & The square root of x \\
& sin(x) & The sine of x \\
& cos(x) & The cosine of x \\
& tan(x) & The tangent of x \\
& atan(x) & The arctangent of x \\
& log(x) & The logarithm of x \\
& exp(x) & The exponential value of x \\
& abs(x) & The absolute value of x \\
\bottomrule
\end{tabular}
\end{table}

\begin{figure*}[t]
  \centering
  \begin{subfigure}[b]{0.7\textwidth}
    \centering
    \includegraphics[width=\linewidth]{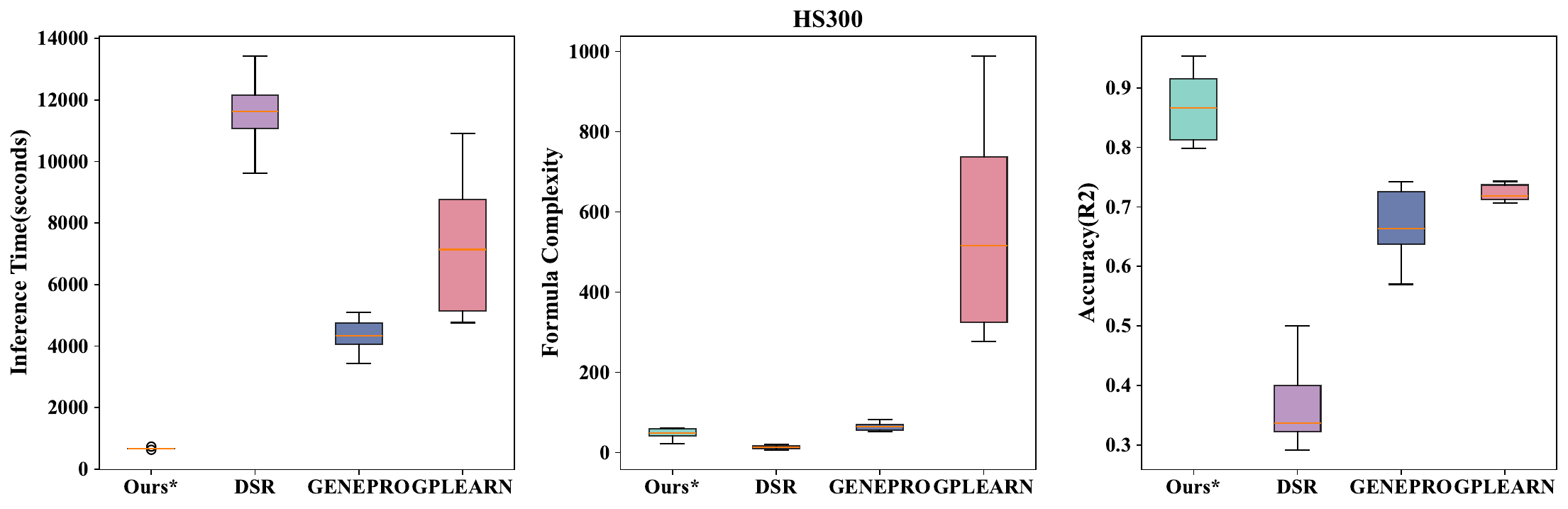}
  \end{subfigure}
  \hfill
  \begin{subfigure}[b]{0.7\textwidth}
    \centering
    \includegraphics[width=\linewidth]{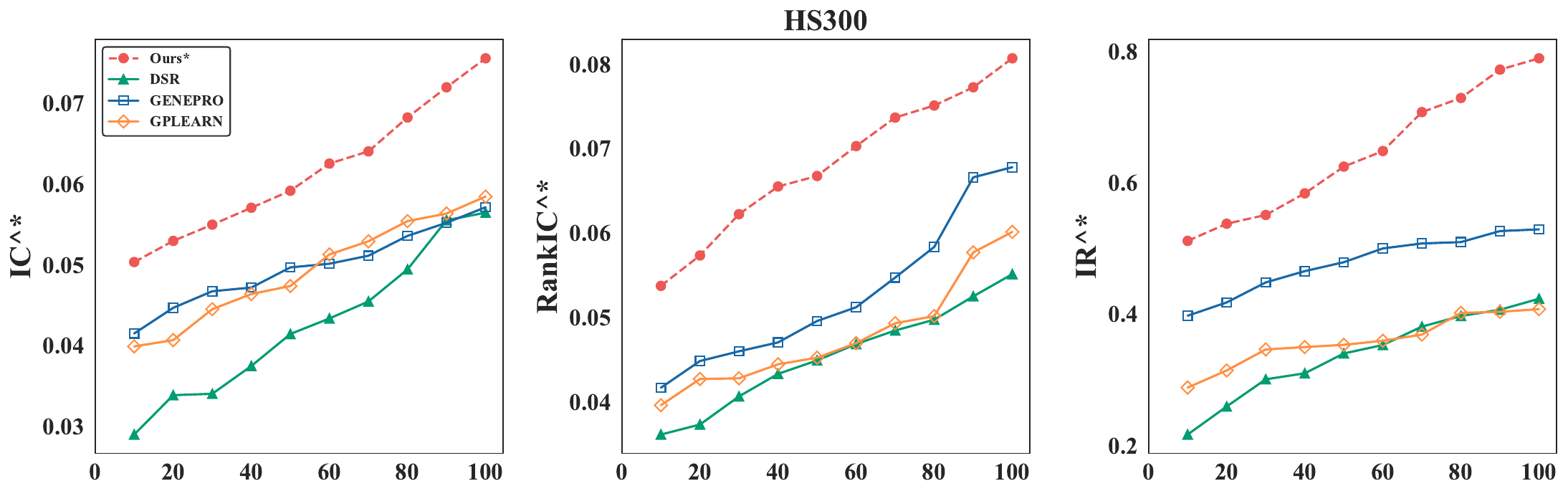}
  \end{subfigure}
  \caption{(a) Performance of different methods on HS300 Index (China market). (b) A comparison of different factor generation methods using $IC^*$, $Rank IC^*$ and $IR^*$ for various factor pool sizes on HS300 Index.}
  \label{fig:lbox_hs300}
\end{figure*}

% \begin{figure}[t]
%   \centering
%   \begin{minipage}{0.36\textwidth} 
%     \includegraphics[width=\linewidth]{heatmap_sp500.pdf}
%   \end{minipage}\\ 
%   \vspace{-0.02\textwidth} 
%   \begin{minipage}{0.3\textwidth} 
%     \includegraphics[width=\linewidth]{AnonymousSubmission/LaTeX/backtest_S&P500.pdf}
%   \end{minipage}
%   \caption{(a) The correlation of the HF risk factor collections on S\&P500 Index. (b) Portfolio simulations on the S\&P500 Index: a backtesting comparison.}
%   \label{fig:heatmap_backtest_sp500} 
% \end{figure}

\begin{figure*}[t]
  \centering
  \begin{minipage}{0.35\textwidth} 
    \includegraphics[width=\linewidth]{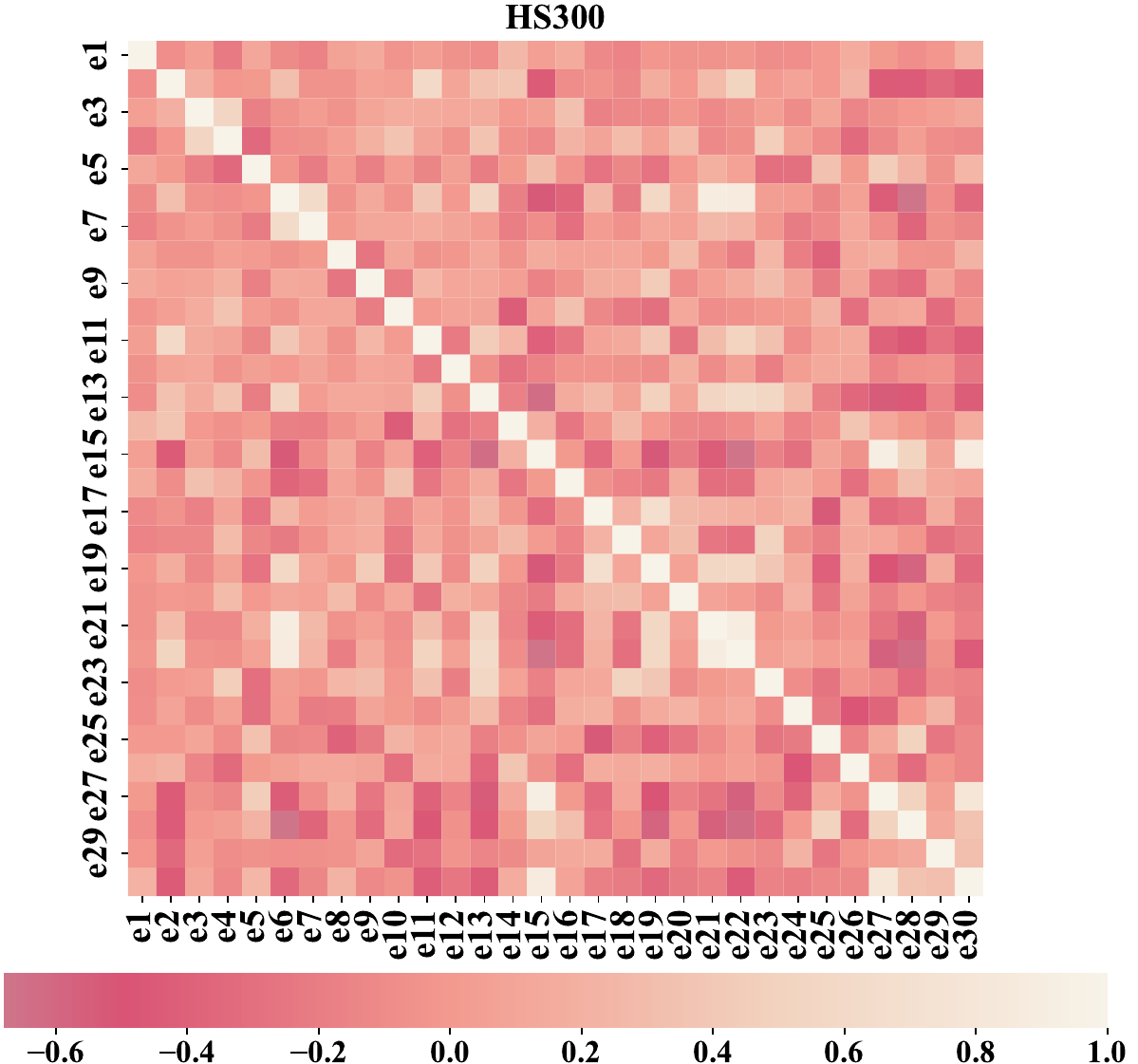}
  \end{minipage}
  \begin{minipage}{0.325\textwidth} 
    \includegraphics[width=\linewidth]{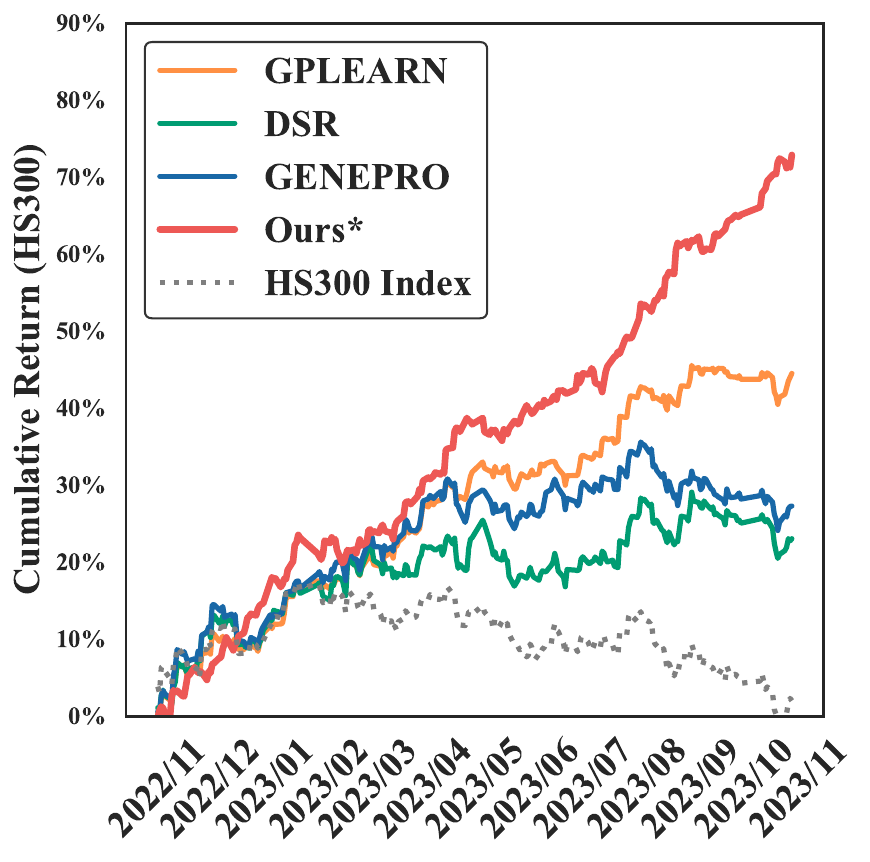}
  \end{minipage}
  \caption{(a) The correlation of the HF risk factor collections on HS300 Index. (b) Portfolio simulations on the HS300 Index: a backtesting comparison.}
  \label{fig:heatmap_hs300}
\end{figure*}

\subsection{Collection of operators} \label{Collection of operators}
Table \ref{tab:operators} lists the unary and binary operators used in this paper. We use these binary operators and 10 unary operators to generate HF risk factor expressions. It should be noted that these operators are based on cross-section, not on time series. Since this paper considers two input options for HFT market data sets of different countries, one of which contains time series features reflecting different time windows, we do not discuss the operators at the time-series level.

\subsection{Performance of different methods under the China HFT market} \label{Performance of different methods under the Chinese high-frequency trading market}
As depicted in Fig. \ref{fig:lbox_hs300}(a), our method demonstrates a significant advantage in terms of inference time, outperforming the slowest method, DSR six times. The concentrated distribution of inference times indicates high stability and efficiency of our method. GPLEARN, while faster than DSR, exhibit more dispersed distributions, suggesting their performance is more susceptible to variations in the stock data set.
In terms of expression complexity, DSR generates the most concise expressions, suggesting its ability to find simple and effective HF risk factor expressions. Regarding prediction performance, our method generates HF risk factor expressions with a value of $R^2$ of 0.87, significantly outperforming other methods. The next best method is GPLEARN, followed by GENEPRO, with DSR performing the worst.

\subsection{Comparison of formulaic generators with varying pool capacity(HS300 Index).} \label{Comparison of formulaic generators with varying pool capacity(SSE50)}
We further extend our analysis to evaluate the performance of different methods in generating HF risk factor collections for the China stock market. As depicted in Fig. \ref{fig:lbox_hs300}(b), our method, IRFT, consistently exhibits the highest values of $IC^*$, $Rank IC^*$, and $IR^*$ across all factor pool sizes. This indicates its superior predictive ability in generating HF risk factors. The performance of the other three methods does not differ significantly.

\subsection{HF risk collections for investment simulation(SSE50 Index).} \label{HF risk collections for investment simulation(SSE50 Index)}

As depicted in Figure \ref{fig:heatmap_hs300}(a), we have implemented a filtering process to delete HF risk factors with high correlation. Consequently, the HF risk factor collection retains 10 unique HF risk factors. As illustrated in Figure \ref{fig:heatmap_hs300}(b), we recorded the cumulative net value of all strategies over a one-year test period in the China HFT market. Our framework outperforms other methods, achieving the highest profit and exceeding other SR methods by 30\%. Next, GPLEARN obtains higher portfolio returns than the remaining two benchmarks. The one that receive the lowest return is DSR. The backtest results demonstrate that our method still has maintained stable growth throughout the backtest period. 

\end{document}